\documentclass[english,a4paper]{JAC2003}
\usepackage[T1]{fontenc}
\usepackage[latin9]{inputenc}
\usepackage{array}
\usepackage{amsmath}
\usepackage{graphicx}
\usepackage{esint}
\makeatletter

\providecommand{\tabularnewline}{\\}

\makeatother

\usepackage{babel}

\usepackage{varwidth}

\usepackage{xcolor}

\begin{document}

\title{BEAM--BEAM STUDY OF ERL BASED eRHIC}

\author{Y. Hao, V.N. Litvinenko, V. Ptitsyn, BNL, Upton, NY, USA}
\maketitle
\begin{abstract}
Beam--beam effects in eRHIC, the proposed ERL-based Electron--Ion Collider
(EIC) at BNL, have several unique features distinguishing them from
those in hadron-colliders and lepton-colliders. Taking the advantage of the
fact that the electron beam is used only once, we expect the luminosity
to be 10 times greater than for the ring--ring collision scheme with similar parameters.
However, without instituting proper treatments, the quality of electron
and hadron beams can undergo degradation or even beam loss, driven
by the beam--beam interactions. We will discuss the harmful effects,
including the disruption and mismatch effect of the electron beam,
the kink instability and the noise heating of the ion beam and the
possible countermeasures.
\end{abstract}

\section{Introduction}

\begin{figure}
\includegraphics[width=1\columnwidth]{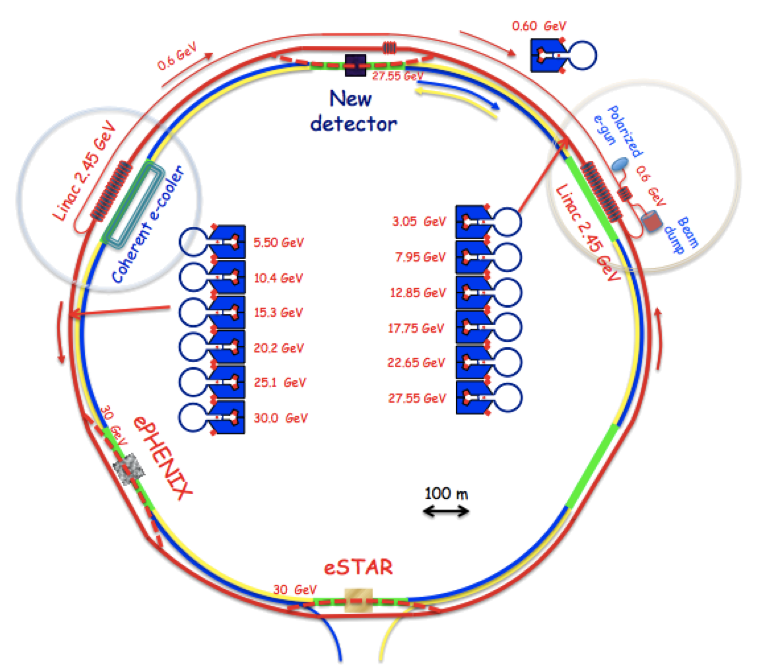}

\caption{\label{fig:eRHIC layout}eRHIC design layout. The blue and yellow
curves represent the existing blue and yellow rings of RHIC. The red
curve illustrates the new ERL accelerator for the electron beam.}
\end{figure}

The main advantage of an energy recovery linac (ERL) based
electron--ion collider (EIC) compared with a ring--ring collider is the higher
achievable luminosity of the former. In an ERL-based EIC, which we
also call a linac-ring scheme, the electron bunch collides only once
with the ion bunch and thereafter is recycled. Hence, the beam--beam
parameter for the electrons in ERL scheme can exceed by a large margin
(as in Table \ref{tab:Parameter_of_eRHIC}) that permissible for electron
circulating in a ring. While the beam--beam parameter for the ions
remains the same in both schemes, the luminosity achieved in the linac-ring
collision scheme exceeds that of the ring--ring collider scheme
between 10 and 100 times \cite{eRHIC_position}. Figure \ref{fig:eRHIC layout}
illustrates the layout of eRHIC, the EIC proposed in Brookhaven National
Laboratory. Table \ref{tab:Parameter_of_eRHIC} lists its design
parameters.

In the new parameter range of eRHIC, the electron beam is subject
to a very strong beam--beam effects that create a new set of beam dynamics
effects. First, the electron beam experiences significant disruption
and mismatch effects due to the beam--beam interaction. Second, the
collective motion of the electron beam inside the ion beam during
their collision can cause a new head--tail type of instability, named
'kink instability'. And the ion beam can be heated up by the noise
of the fresh electron beam each turn. In this paper, we will report
our studies on those individual effects and carry out countermeasures
to the harmful ones.

\begin{table}
\caption{Parameter range of eRHIC\label{tab:Parameter_of_eRHIC}}

\begin{tabular}{|>{\centering}m{0.65\columnwidth}|>{\centering}m{0.25\columnwidth}|}
\hline
Parameters & Range\tabularnewline
\hline
\hline
Electron beam energy (GeV) & 5-30\tabularnewline
\hline
Ion beam energy (GeV) (proton)  & 50-250\tabularnewline
\hline
Electron beam disruption parameter & 5-142\tabularnewline
\hline
Ion beam--beam parameter & 0.015\tabularnewline
\hline
Ion bunch length (cm) & 8.3\tabularnewline
\hline
Electron bunch length (cm) & 0.2-0.4\tabularnewline
\hline
Electron and ion $\beta^{*}$ (cm) & 5\tabularnewline
\hline
Ion synchrotron tune & 0.004\tabularnewline
\hline
\end{tabular}
\end{table}

\section{Electron disruption effects}

The electron beam experiences very strong beam--beam force from the
ion beam in the interaction region. The force will make the electron
beam oscillate inside the ion beam and deform the distribution of
the electron beam, as studied in \cite{prstab}. We found that the
disruption parameter $d_{e}=l_{i,z}/f_{e}$ is convenient to characterize
the oscillation of the electron beam, where $l_{i,z}$ is the ion
bunch length and $f_{e}$ is the focal length of the linearized beam--beam
interaction. For an ion beam with Gaussian longitudinal distribution,
the number of oscillations $n$ of the electron beam inside the ion
beam is
\[
n=\frac{\sqrt{d_{e}}}{\left(2\pi\right)^{3/4}}\approx\frac{\sqrt{d_{e}}}{4}.
\]
Thus, for the eRHIC parameters, a single electron will oscillates
up to 3 full oscillations in the ion beam.

\begin{figure}
\includegraphics[width=1\columnwidth]{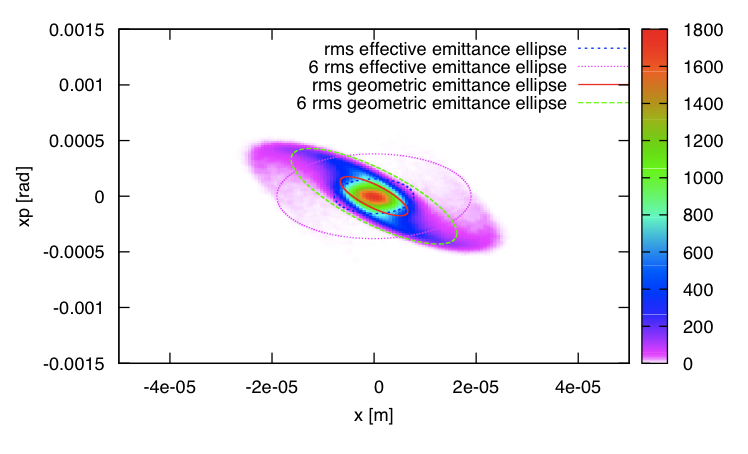}

\includegraphics[width=1\columnwidth]{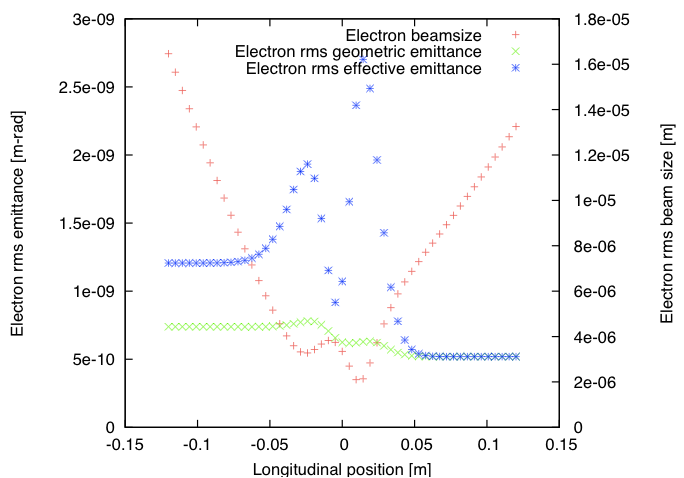}

\caption{\label{fig:de27s}The top figure shows the phase space distribution of the electron beam after
collision, and bottom figure shows the evolution of the electron beam size and emittance,
for $d_{e}=27$. In the top figure, the r.m.s.\ and 6 r.m.s.\  ellipses
for both geometric and effective emittance, respectively, are plotted.}
\end{figure}

\begin{figure}
\includegraphics[width=1\columnwidth]{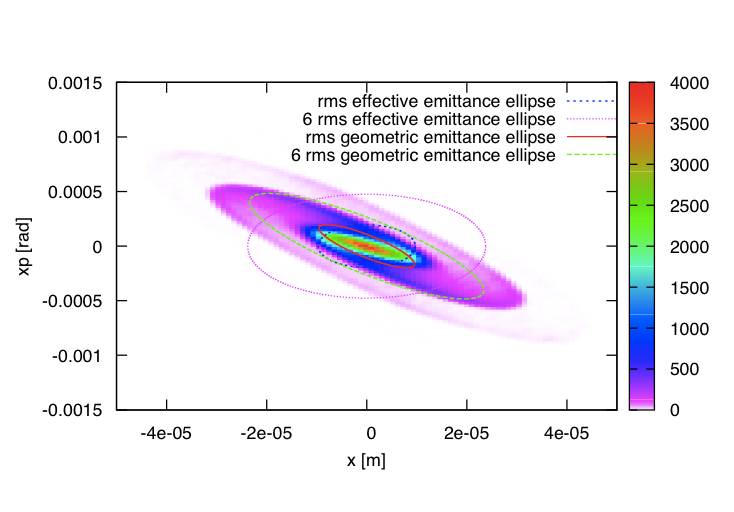}

\includegraphics[width=1\columnwidth]{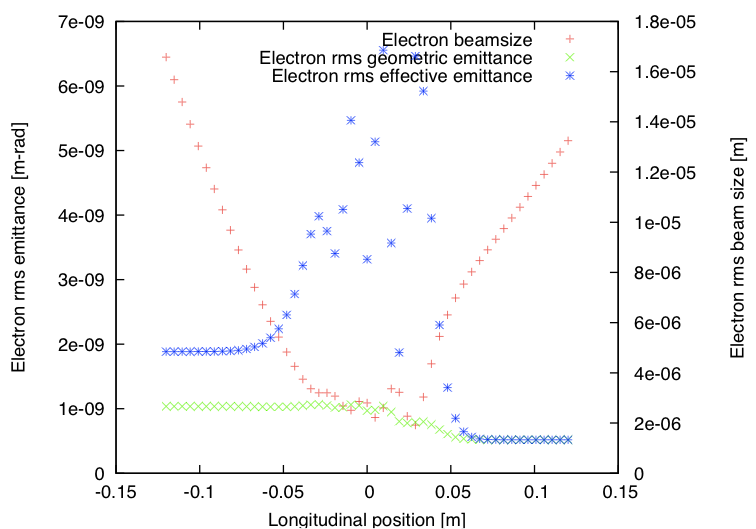}

\caption{\label{fig:de150s}The top figure shows the phase space distribution of the electron beam after
collision, and bottom figure shows the evolution of the electron beam size and emittance,
for $d_{e}=150$. In the top figure, the r.m.s.\ and 6r.m.s.\  ellipses
for both geometric and effective emittance, respectively, are plotted.}
\end{figure}

We use simulation code, EPIC \cite{thesis}, to calculate the electron
beam evolution inside the opposing ion beam. Figure~\ref{fig:de27s}
and Fig.~\ref{fig:de150s} illustrate the examples of the electron beam
distribution after the collision and the e-beam evolution inside the
ion beam. The former correspond to the case of $d_{e}=27$, and latter
for $d_{e}=150$. In the electron beam distribution plots, the nonlinear
force deform its initial Gaussian distribution completely. The electrons
with larger betatron amplitude rotate slower than those in the core.
Therefore the distribution after collision forms a spiral shape. We
use 2 different definitions of beam emittance to characterize the
occupied phase space area. One is the r.m.s.\ geometric emittance obtained
from the beam distribution, written as
\begin{equation}
\varepsilon_{x}=\sqrt{\left\langle {\left({x-\bar{x}}\right)^{2}}\right\rangle \left\langle {\left({x'-\bar{x}'}\right)^{2}}\right\rangle -\left\langle {\left({x-\bar{x}}\right)\left({x'-\bar{x}'}\right)}\right\rangle ^{2}}.\label{eq:Definition of Geometric Emittance}
\end{equation}
The other emittance uses the design optics function 
and is called effective emittance. It is defined as the half of the
average value of the Courant--Snyder invariant of all macro-particles based
on the design lattice
\begin{equation}
C\left({\tilde{x},\tilde{x}'}\right)=\gamma\tilde{x}^{2}+2\alpha\tilde{x}\tilde{x}'+\beta\tilde{x}'^{2}.
\end{equation}

\begin{figure}
\includegraphics[width=1\columnwidth]{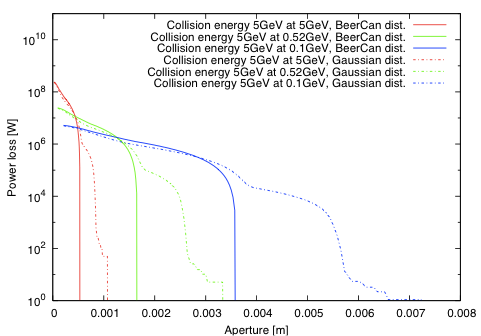}

\caption{\label{fig:aperture requirement}The aperture requirement is shown of the energy
recovery pass downstream IP. A maximum 10~m $\beta^{*}$ is assumed
in all arcs.}
\end{figure}

In the e-beam distribution plots of Fig.~\ref{fig:de27s} and Fig.~\ref{fig:de150s},
both emittances are represented as ellipses of 1~r.m.s.\  value and
6~r.m.s.\ value. The evolution plots illustrate the evolution of the 
2~r.m.s.\ emittance and the r.m.s.\  beam size. These plots clearly show the
mismatch between the beam distribution and the design optics due to
the beam--beam interaction. The effective emittance will determine
the aperture requirement of the magnet downstream of interaction point
(IP), as shown in Fig.~\ref{fig:aperture requirement}. The calculated
aperture shows that the small-gap magnet designed for eRHIC is suitable
for the ERL energy recovery passes.

\section{Kink instability and its mitigation methods}

\begin{figure}
\includegraphics[width=1\columnwidth]{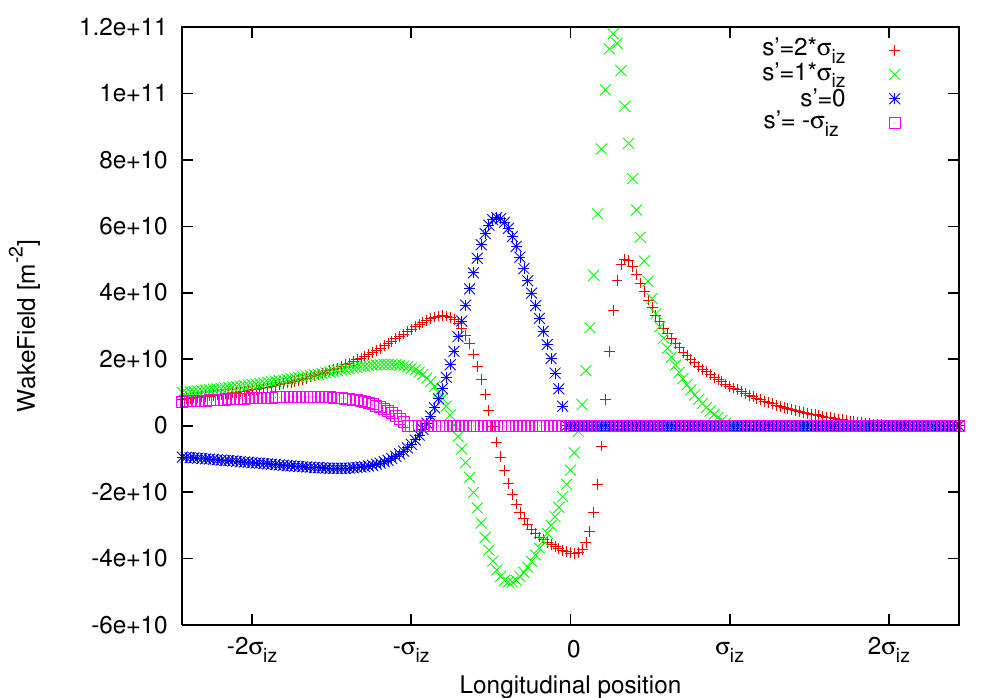}

\caption{\label{fig:kink-wake}The example of the kink wake field with the
beam--beam parameter of the ion beam $\xi_{p}=0.015$ is shown. The electron
beam has disruption parameter $d_{e}=27$. In the figure, the electron
beam travels from the positive $s$ to negative. The symbol $s^{'}$ denotes
the location of the introduced offset.\label{fig:examples-of-wake}}
\end{figure}

The kink instability develops due to the electron beam passes the
imperfection of the head of the ion beam to its tail. Therefore, for
the ion beam, the beam--beam interaction behaves as an effective wake
field. If we assume both beams have only infinitesimal offsets, the
wake field

\begin{equation}
W\left(s,s^{\prime}\right)=\dfrac{\gamma_{i}}{Z^{2}N_{ib}r_{i}}\dfrac{\Delta x^{\prime}\left(s\right)}{\Delta x\left(s'\right)}
\end{equation}
can be retrieved from simulation, where $N_{ib}$ is the number of
ions in the slice, $\gamma_{i}$ is the energy of the ion beam and
$r_{i}$ is the classical radius of the ion beam. The wake field is
illustrated in Fig.~\ref{fig:kink-wake}.

\begin{figure}
\includegraphics[width=1\columnwidth]{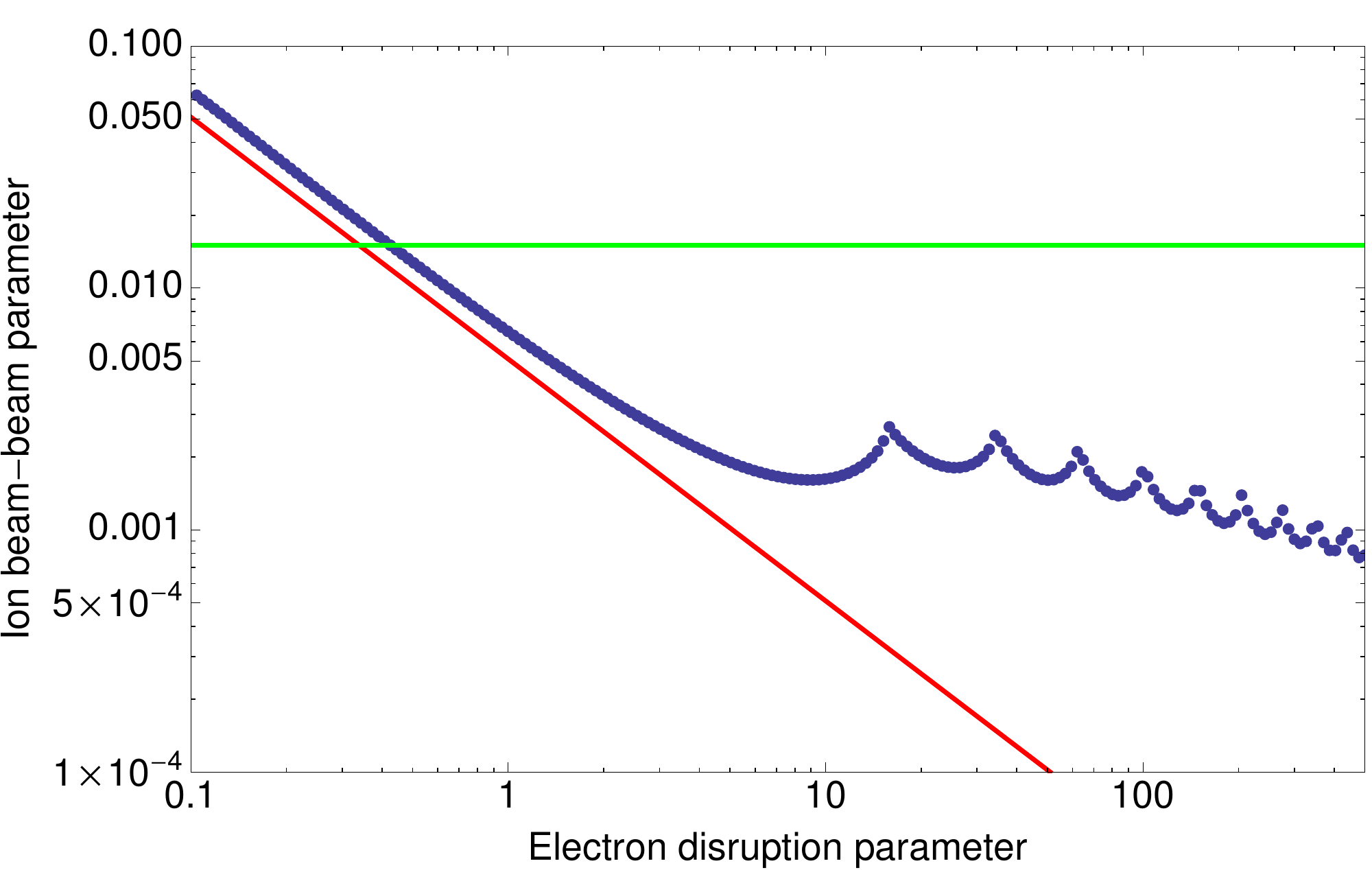}

\caption{\label{fig:kink-threshold}The threshold of kink instability, with
the choice of the synchrotron tune 0.004, is shown. The blue dots denote the
threshold calculated from the 51 macro-particles circulant matrix method.
The red line represents the simple threshold form from equation $\xi_{i}d_{e}<4\nu_{s}/\pi$.
The green line corresponds to $\xi_{i}=0.015$, which is the design beam--beam
parameter of ion beam in eRHIC. }
\end{figure}

The threshold of the strong head--tail instability (the kink instability)
can be calculated using the 2-particle model or the multi-particle
model\cite{PAC11_kink}. Both models are based on linearized beam--beam
forces. For a 2-particle model, the threshold is simply: $\xi_{i}d_{e}<4\nu_{s}/\pi$.
However, to model the electron beam correctly in the high disruption parameter
case, the multi-particle model should be used, predicting the threshold
as in Fig.~\ref{fig:kink-threshold}.

\begin{figure}
\includegraphics[width=1\columnwidth]{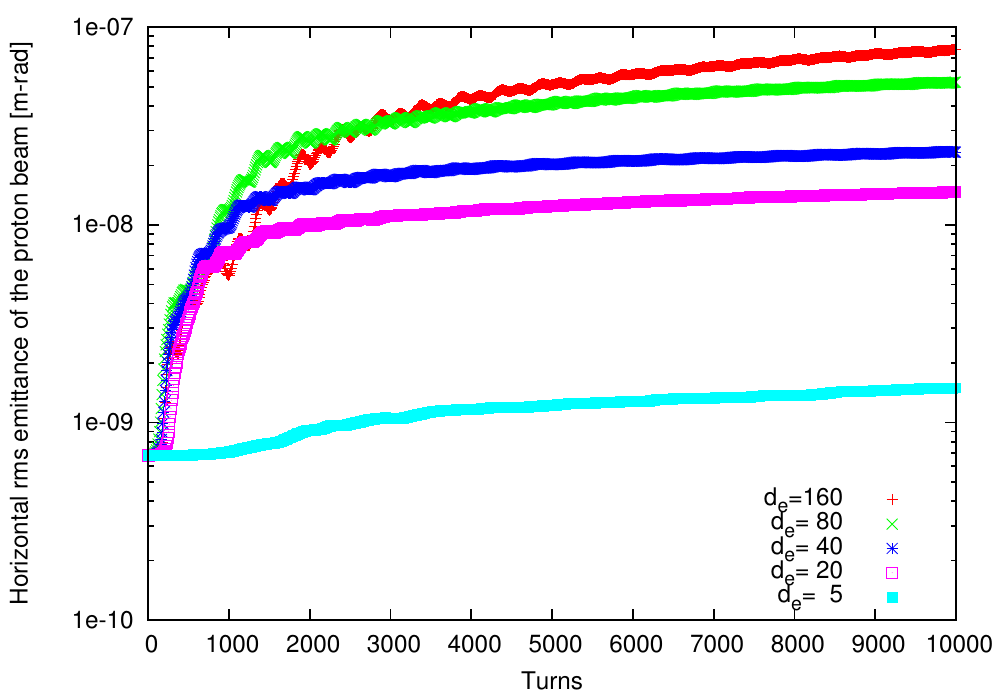}

\includegraphics[width=1\columnwidth]{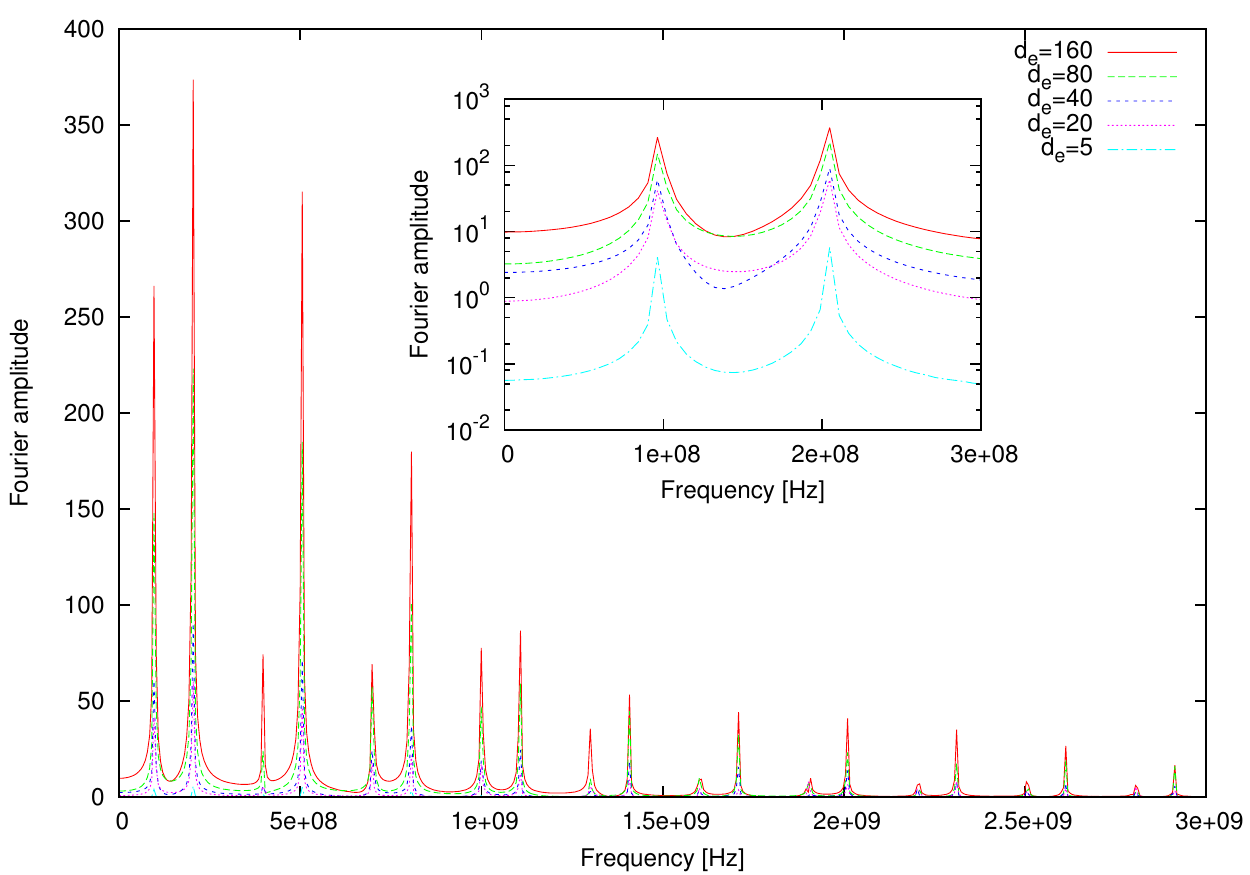}

\caption{\label{fig:kink-simulation}The top figure shows the proton beam emittance growth
due to the kink instability at different disruption parameters with
the chromaticity of both transverse directions set at $+2$ units,
and the beam--beam parameter of the proton beam at 0.015. The bottom figure shows
Fourier spectrum of the turn by turn proton slice centroid data. The
proton beam is cut to 100 longitudinal slices for this calculation.}
\end{figure}

Both linear models predict that the parameter of the eRHIC exceeds
the threshold. A simulation using nonlinear beam--beam forces is required
to confirm this understanding. Figure \ref{fig:kink-simulation} shows
the emittance growth associated with the kink instability at different
disruption parameters of the electron beam. Even with the lowest disruption
parameter, $d_{e}=5$, the system is not stable at +2 chromaticity
(the nominal value of RHIC operation), although the emittance growth
in this case is much less than those with higher $d_{e}$. If we increase
the chromaticity to stabilize the emittance growth, it requires unreasonable
values. Therefore, a dedicate feedback system is desired as a countermeasure.

\begin{figure}
\includegraphics[width=1\columnwidth]{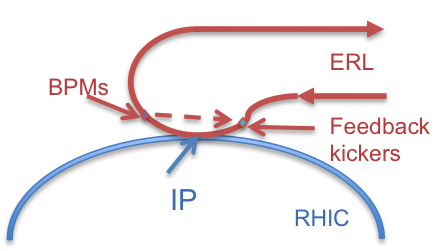}

\caption{\label{fig:kink-feedback-layout}The schematic layout is shown of the feedback
system I for mitigating the kink instability in eRHIC.}
\end{figure}

The first feedback system~\cite{PAC11_kink}, shown in Fig.~\ref{fig:kink-feedback-layout},
takes full advantage of flexibility of a linac-ring scheme, which
has the following procedures. We steer the fresh electron bunch before
collision based on the transverse offset of the last-used electron
bunch that collides with the same ion bunch. Then the centroid of
the new electron bunch will oscillate within the opposing ion bunch
due to the focusing beam--beam force. We are expecting that oscillation
of the centroid of the electron bunch gives the ion bunch proper kicks
to correct the offset of the ion bunch before the visible adverse
effect, such as emittance growth and luminosity loss, due to the kink
instability.

Mathematically, we introduce the offset by modifying the motion of
the centroid of the electron bunch based on the information from the
last one:
\begin{equation}
\binom{\bar{x}_{e}}{\bar{x}_{e}^{\prime}}_{n+1,i}=M_{f}\binom{\bar{x}_{e}}{\bar{x}_{e}^{\prime}}_{n,f}.\label{eq:feedback_matrix}
\end{equation}
Here, the subscript $n$ denotes the electron--ion collision in $n^{th}$
turn, and the subscripts $i$ and $f$ respectively represent the bunch
centroid before and after collision. Map $M_{f}$ defines the algorithm
of the feedback system. Here, for
simplicity and easier realization, we limit $M_{f}$ to be a matrix.

\begin{figure}
\includegraphics[width=1\columnwidth]{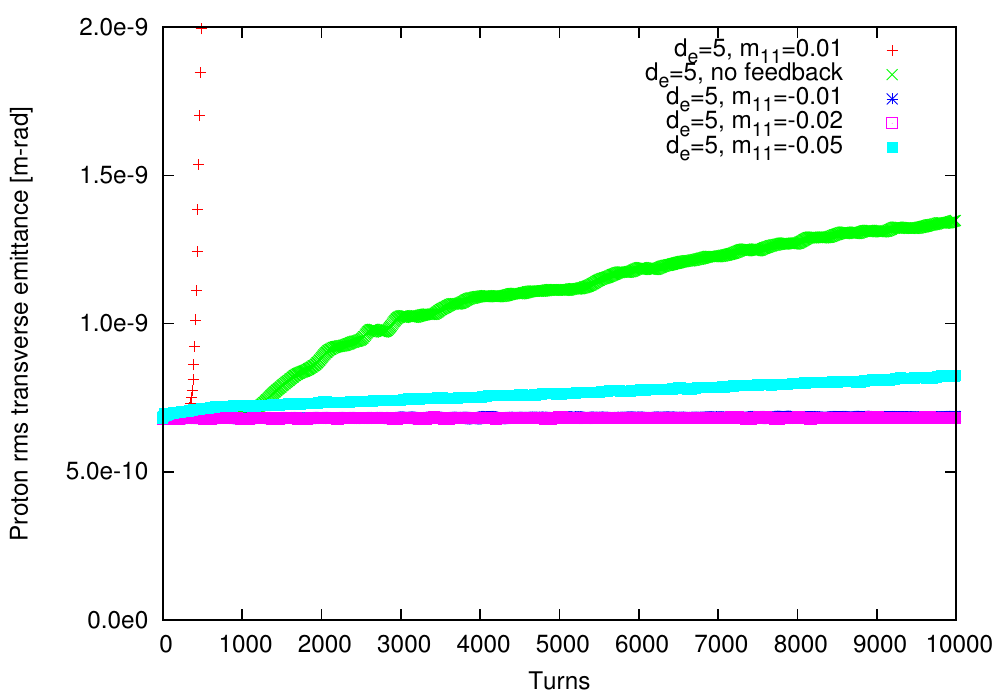}

\caption{\label{fig:kink-feedback1-de5}The effect of the feedback system at
disruption parameter 5 is shown.}
\end{figure}

\begin{figure}
\includegraphics[width=1\columnwidth]{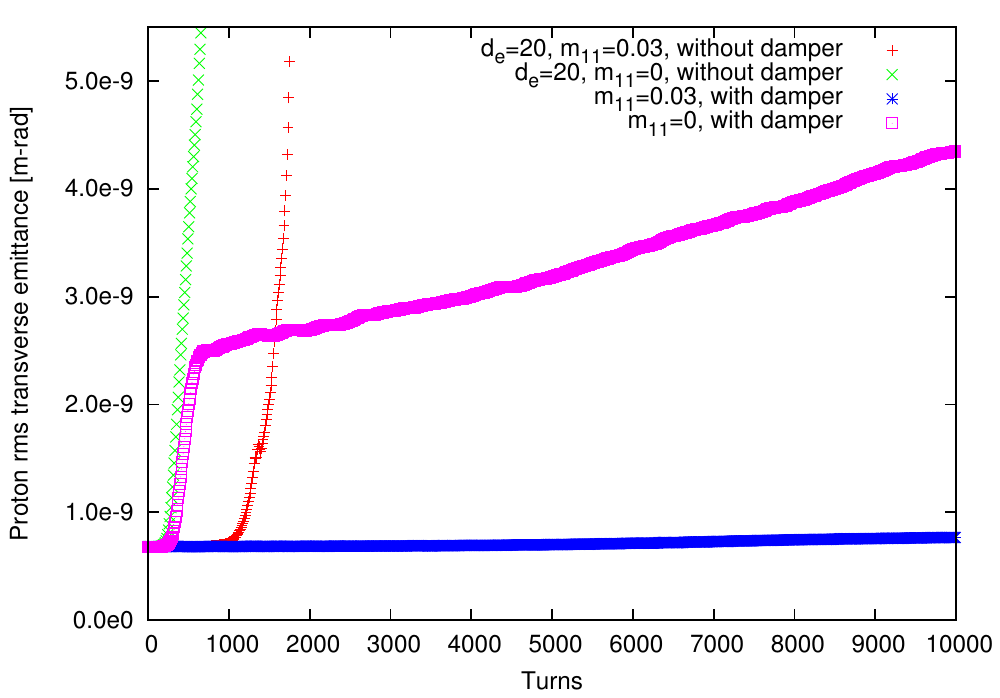}

\caption{\label{fig:kink-feedback1-de20}The effect of the feedback system
at disruption parameter 20 is shown.}
\end{figure}

Figure \ref{fig:kink-feedback1-de5} shows the effect of this feedback
system at disruption parameter 5. In this case, the emittance growth
due to the kink instability is suppressed with proper amplitude of
the feedback gain $m_{11}$ (-0.01 or -0.02) without a noticeable
decreasing in luminosity. An incorrect sign of the gain may boost
the instability, as shown by the red dots in Fig.~\ref{fig:kink-feedback1-de5}.

When the disruption parameter exceeds 15, this feedback system itself
can not stabilize the emittance, because the system will excite the
instability of the rigid mode while it can correct the head--tail mode
of the ion beam. Therefore we have to add the transverse bunch-by-bunch
damper to damp the rigid mode of the ion beam simultaneously. The
result for $d_{e}=20$, as an example, is shown in Fig.~\ref{fig:kink-feedback1-de20}.
The red dots show the case with the feedback gain of $m_{11}=0.03$
without transverse damper. The centroid of the ion bunch becomes unstable
and causes fast emittance growth due to the offset of two beams. By
applying the bunch-by-bunch feedback in the simulation, the ion centroid
is stable and the emittance growth is prevented (blue curve).

The simple feedback loses its efficiency when $d_{e}>25$. In this
range, the electron beam oscillate too fast and the frequency of the
oscillation does not match that of the lowest instability mode. We
need an alternative feedback scheme for this disruption parameter range,
such as a traditional pick-up and kicker system in the ion beam, to
suppress the instability coherently \cite{PAC12_kink}.

For the pickup--kicker system, the effect can also be modelled as a
wake field. If we assume the system has a uniform frequency response
with low and high frequency limits $f_{L}$ and $f_{H}$, the corresponding
wake field of this system is \cite{cooling}

\begin{equation}
W\left(\tau\right)=R\int_{f_{L}}^{f_{H}}\cos\left(2\pi f\tau\right)\,df,
\end{equation}
where $R$ is related to gain of the amplifier between the pickup
and the kicker.

We fix the low frequency limit to 50 MHz, which is below the first
peak in the bottom figure of Fig.~\ref{fig:kink-simulation}. Then we vary
the high frequency limit to find the requirement for the individual disruption
parameter.

\begin{figure}
\includegraphics[width=1\columnwidth]{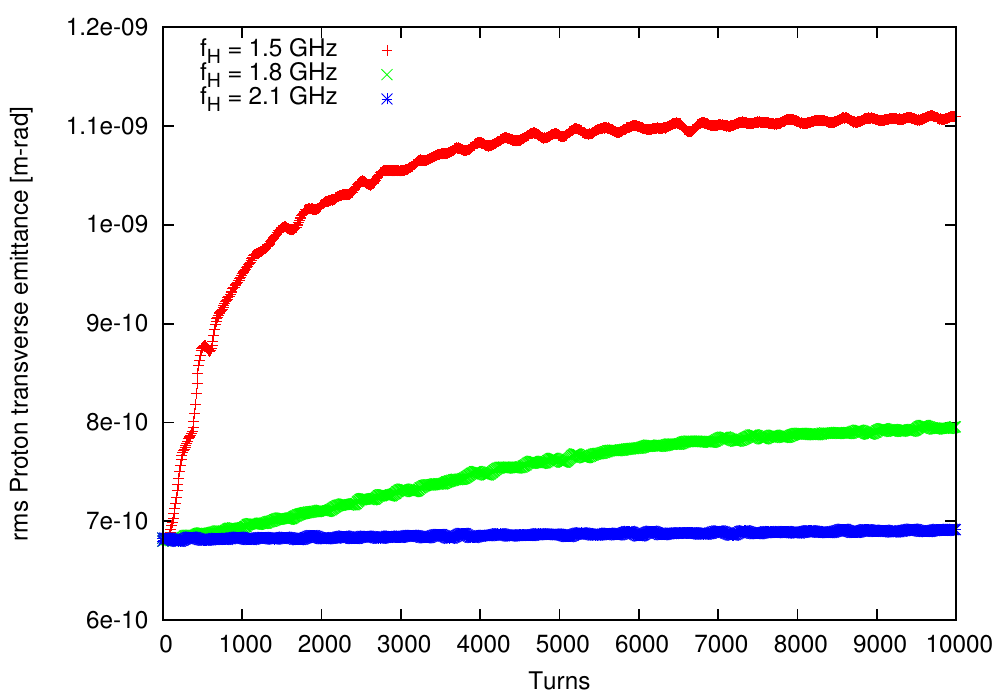}

\caption{\label{fig:kink-feedback2-de150}The comparison is shown of kink instability dampings
with different high frequency limits $f_{H}$ when the disruption parameter
$d_{e}=150$. The gain of the feedback is selected to minimize the
emittance growth ion beam. }
\end{figure}

\begin{figure}
\includegraphics[width=1\columnwidth]{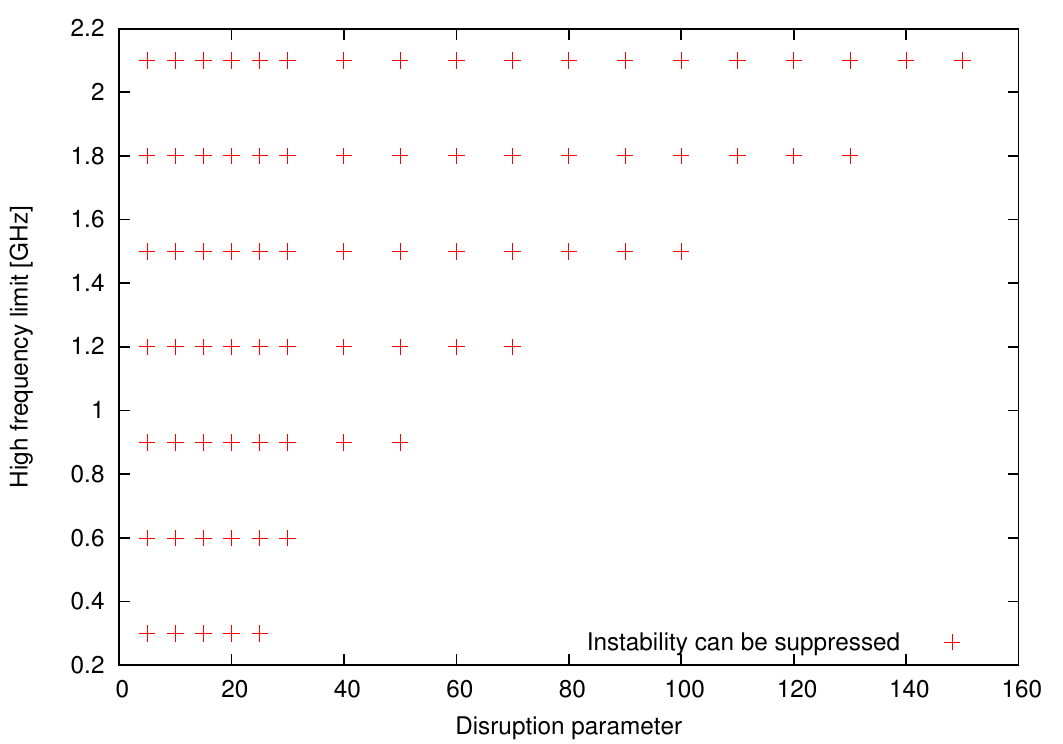}

\caption{\label{fig:kink-feedback2-des}The relation is shown between the required high
frequency limit $f_{H}$ and the electron disruption parameter $d_{e}$.
Each point shows where the instability can be suppressed in the corresponding
parameter ($f_{H}$ and $d_{e}$) with proper amplitude. For all calculations,
the low frequency limit is set at 50~MHz }
\end{figure}

Figure \ref{fig:kink-feedback2-de150} shows that the required $f_{H}$
is at least 2.1 GHz to suppress the kink instability when $d_{e}$
is 150. For other $d_{e}$, as shown in Fig.~\ref{fig:kink-feedback2-des},
the required $f_{H}$ is a monotonically increasing function of $d_{e}$.
Therefore, we demonstrated that the kink instability will be suppressed
by a pickup and kicker system with whole electron beam disruption
parameter range (5--150), if the required frequency bandwidth is selected.

\section{Noise heating effect of the ion beam}

Since the ion beam always collides with fresh electron bunches, the
electron beam parameter fluctuation will affect the ion beam through
the beam--beam interaction. The fluctuations can be classified as two
types. The first is dipole errors due to the electron beam transverse
position offset; the second is quadrupole error due to the fluctuation
of the electron beam intensity or transverse beam size.

If the noise of the electron beam is considered as white noise,
i.e.\ a uniform spectrum in frequency domain, the effect of both dipole error
and quadrupole errors can be evaluated analytically. For the quadrupole
errors, the r.m.s.\  beam size of the ion beam is expected to grow exponentially,
with the rising time
\[
\tau=\frac{T}{4\pi^{2}\xi_{i}^{2}}\frac{1}{\left(\delta f/f\right)^{2}},
\]
where $\xi_{i}$ is the beam--beam parameter of the ion beam, $T$
is the revolution period and $\delta f/f$ is the r.m.s.\  error of the
beam--beam focal length. For eRHIC parameters, to get a slow rising time
(\textasciitilde{}10 hours), the relative error of the electron beam parameter
should be better than $2\times10^{-4}$. 
A Lorentz frequency spectrum $g(\omega)=1/(\omega^{2}+\alpha^{2}\omega_{0}^{2})$
is considered, where $\alpha$ is a free parameter much less than
1 and $\omega_{0}$ is the angular revolution frequency of the ion
ring. In this case, the rising time $\tau$ is lengthened to $\tau/R(\alpha)$,
where

\begin{eqnarray*}
R(\alpha) & = & \frac{1-\exp(-2\alpha)}{1+\exp(-2\alpha)-2\cos(4\pi\nu)\exp(-\alpha)}\\
 & = & \frac{\alpha}{1-\cos(4\pi\nu)}+O(\alpha^{3}).
\end{eqnarray*}

For the dipole errors, the ion beam is kicked turn by turn due to
the electron beam random offset. By following the well-known random
walk formulas, the ion beam displacement gives $\sqrt{<x_{i}^{2}>(t)}=\sqrt{t/\tau+<x_{i}^{2}>(0)}$
and $1/\tau=8\pi^{2}\xi_{i}^{2}<d_{n}^{2}>/T$, where $d_{n}$ is
the $n^{th}$ turn electron beam displacement at IP. We need a
bunch-by-bunch transverse damper in the ion ring to compensate the dipole
heating up effect.

\section{Conclusion}

We report on the key finding for distinct beam--beam effects in the
ERL based eRHIC. Our study identified the challenges as well as possible
countermeasures for both the electron and the ion beams.

A dedicated feedback system is required to suppress the emittance
growth caused by the kink instability. We proposed two possible feedback
systems. The feedback applied to the electron beam works for moderate
values of the disruption parameter, e.g.\ $d_{e}<25$. A traditional broad-band
pickup and kicker feedback system can damp the instability for the
whole range of the disruption parameter expected in eRHIC.

We report on the requirement for the intensity and beam size stability
of the electron beam to avoid the hadron beam emittance growth caused
by noise in beam--beam interactions. We also established a need for
a transverse bunch-by-bunch damper to compensate for the possible heating
effect caused by random noise in the transverse displacement in the electron
beam.


\begin{thebibliography}{9}
\bibitem{eRHIC_position}
V. Ptitsyn \emph{et al.}, eRHIC Accelerator Position Paper, Tech. Rep.
(C-AD, BNL, 2007).

\bibitem{prstab}
Y. Hao and V. Ptitsyn, \emph{Phys. Rev. ST Accel. Beams} \textbf{13} (2010) 071003.

\bibitem{thesis}
Y. Hao, Beam-Beam Interaction Study in ERL based eR- HIC, Ph.D. thesis,
Indiana University, 2008.

\bibitem{PAC11_kink}
Y. Hao, V. N. Litvinenko and V. Ptitsyn, \emph{Phys. Rev. ST Accel. and Beams} \textbf{16} (2013) 101001.

\bibitem{PAC12_kink}
Y. Hao \emph{et al.}, Kink Instability Suppression with Stochastic Cooling
Pickup and Kicker, Proc. Int. Particle Accelerator
Conf., New Orleans, NM,  2012.

\bibitem{cooling}
M. Blaskiewicz and J.M. Brennan, WEM2105, Proc.COOL,
Bad Kreuznach, 2007.

\end{thebibliography}
\end{document}